\documentclass[%
 reprint,
 amsmath,amssymb,
 aps,
]{revtex4-2}
\newcommand{\SFT}{\rm SFT}
\newcommand{\WFT}{\rm WFT}

\newcommand{\SBFT}{\rm SBFS-FT}
\newcommand{\TS}{\rm TS}
\usepackage{changes}
\usepackage{graphicx}
\usepackage{dcolumn}
\usepackage{bm}

\begin{document}

\preprint{APS/123-QED}

\title{Symmetry-Based Parameter Efficient Low-Rank Fine-Tuning for \\ High-Throughput Calculation of Lattice Thermal Conductivity }

\author{Rustam Arabov}
\affiliation{Skolkovo Institute of Science and Technology, Bolshoy Boulevard, 30, p.1, Moscow, Russia, 121205}
 
\author{Konstantin Sozykin}%
\affiliation{Skolkovo Institute of Science and Technology, Bolshoy Boulevard, 30, p.1, Moscow, Russia, 121205}

\author{Alexander Shapeev}
\affiliation{Skolkovo Institute of Science and Technology, Bolshoy Boulevard, 30, p.1, Moscow, Russia, 121205}
\affiliation{Digital compounds LLC, Medovaya st., 3, Zarechye, Moscow Region, Russia, 143085}
 
\author{Nikita Rybin}
\affiliation{Skolkovo Institute of Science and Technology, Bolshoy Boulevard, 30, p.1, Moscow, Russia, 121205}
\affiliation{Digital compounds LLC, Medovaya st., 3, Zarechye, Moscow Region, Russia, 143085}

\date{\today}

\begin{abstract}

Accurate and efficient prediction of lattice thermal conductivity (LTC) is essential for developing materials for thermoelectric, microelectronic, and photovoltaic applications. Here, we present a methodology for high-throughput calculation of LTC across large datasets of compounds sharing a common crystal symmetry. Our approach combines fine-tuning of the MatterSim 1M universal machine-learned interatomic potential with LTC calculations based on the solution of the Boltzmann transport equation. We introduce a symmetry-based few-shot fine-tuning strategy (SBFS-FT) based on low-rank adaptation. In this approach, MatterSim 1M is fine-tuned on a small, carefully selected set of compounds that share a common crystal symmetry. The resulting potential is then applicable for accurate LTC calculations for compounds with this crystal symmetry, even if they were not included in the fine-tuning set. We tested our methodology on datasets of half-Heusler compounds (space group 216), materials with space group 225, and spinels (space group 227). We demonstrated that LTC calculations with MatterSim 1M fine-tuned using SBFS-FT exhibited a mean absolute relative difference from the reference results (obtained using moment tensor potentials and polyMLPs) 2–3 times smaller than that achieved with the non-fine-tuned MatterSim 1M potential. Moreover, the accuracy of LTC calculations with MatterSim 1M after SBFS-FT was comparable to that achieved using the MatterSim 1M potentials individually fine-tuned for each compound. Importantly, SBFS-FT enabled generation of the potential suitable for accurate LTC calculations across large compound datasets substantially faster than fine-tuning of MatterSim 1M for each compound individually, reducing the computational time by approximately a factor of 5–7 for a dataset of 100 compounds.

\end{abstract}


\maketitle


\section{\label{sec:intro}Introduction}

The accurate calculation of lattice thermal conductivity (LTC) is crucial for the development of novel materials for thermal management, thermoelectrics, and microelectronics. Traditionally, the forces and energies of interatomic interactions required for LTC calculations are obtained with DFT. The DFT-based approaches are accurate but extremely computationally expensive. The computational cost limits practical applicability of these \textit{ab initio} approaches for high-throughput LTC screening. Use of machine-learned interatomic potentials (MLIPs), such as moment tensor potential (MTP)~\cite{shapeev2016moment} and polyMLP~\cite{polymlp_repository}, within the LTC calculation procedures reduces computational costs without loss if accuracy, as demonstrated in several studies~\cite{mtp_proofs_1, mtp_proofs_2, mtp_proofs_3, mtp_proofs_4, polymlp_ltc, togo2023mdr}. However, a significant limitation is that for such calculations MTPs and polyMLPs must be trained from scratch on the DFT data for each specific compound. Therefore, time- and resource-consuming DFT calculations are also a necessary part of the MLIP-based approaches for obtaining LTC.

To overcome this limitation, the universal MLIPs such as MatterSim~\cite{mattersim} can be applied. These models are pre-trained on large datasets of compounds and can predict the forces and energies of interatomic interactions for any material out-of-the-box. However, these universal potentials often suffer from the potential energy surface (PES) softening effect~\cite{softening}. This effect reduces the accuracy of computing the derivatives of PES with respect to atomic displacements. Therefore, all properties connected to the PES derivatives, including LTC, are predicted incorrectly. This problem can be solved by additional training (fine-tuning) of the universal MLIP on DFT data for each specific compound for which we need to obtain LTC. However, in the case of high-throughput LTC screening, performing this fine-tuning process individually for each material from a large dataset remains a time-consuming process.

To address this challenge, we present a novel methodology for accurate and efficient high-throughput LTC calculations. This methodology allows calculating LTC for large datasets of the compounds with common crystal lattice symmetry. In our approach, MatterSim 1M potential is adapted for a large set of compounds with common symmetry by fine-tuning it on a specifically selected subset of compounds from this set. This approach, referred to as symmetry-based few-shot fine-tuning (SBFS-FT), is significantly faster than fine-tuning of MatterSim 1M for each compound in the dataset followed by LTC calculation. The performance of our LTC computation methodology was tested on the HH130 dataset of Half-Heusler compounds~\cite{hh130}, the 225 dataset (dataset of the compounds with space group 225), and the dataset of spinels extracted from the MDR database~\cite{togo2023mdr}. During LTC calculations for 28 materials from all three datasets, the time taken by SBFS-FT was about 1.5-2 times less than that taken by specific fine-tuning. For LTC calculations the method based on the solution of Boltzmann transport equation (BTE-based method)~\cite{ltc_bte} was used. The mean absolute relative difference between LTC values calculated using the MLIP (MTP or polyMLP) and MatterSim 1M served as a metric for evaluating the accuracy of LTC calculation with the universal MLIP. The SBFS‑FT approach of fine‑tuning MatterSim 1M yielded LTC prediction accuracy comparable to that of MatterSim 1M fine‑tuned for each compound individually, and 2–3 times higher than that of MatterSim 1M without fine‑tuning. Therefore, the SBFS-FT procedure described in the present work is a computationally efficient method of increasing the accuracy of MatterSim in high-throughput LTC calculations with the BTE-based method.

\section{\label{sec:com-met}Computational methodology}

\subsection{\label{sec:datasets}The datasets of compounds}

Three datasets were used to evaluate our LTC calculation methodology. Two of them were obtained from existing sources: the HH130 dataset of half-Heusler compounds~\cite{hh130} and the MDR lattice thermal conductivity calculation database~\cite{togo2023mdr}, from which we selected spinels. The HH130 dataset contains 130 half-Heusler compounds. The MDR database includes 45 spinels among its 4811 compounds. Both datasets provide crystal structures, interatomic potentials (MTPs for HH130 and polyMLPs for MDR), and corresponding training sets with interatomic energies and forces computed using density functional theory (DFT). The training sets contain approximately 220 configurations per compound for HH130 and 200 configurations per compound for MDR. The MDR database also provides LTC values calculated using the polyMLP potentials.

The third dataset was generated in this work and comprised 28 compounds with space group 225, for which NVT molecular dynamics (MD) simulations were performed using a time step of 1 fs. Interatomic energies and forces along the trajectories were initially computed using DFT with the FHI-aims software package~\cite{fhiaims}. For several compounds, MTPs were trained on carefully selected subsets of trajectory frames, as detailed in the Supplementary Material. To ensure consistency with the data used to fit the pre-trained MatterSim potential, interatomic energies and forces for all configurations included in the training sets were recalculated using VASP. The resulting sets, containing approximately 220 configurations per compound, were then used to fine-tune the MatterSim 1M potential for LTC calculations.

Each dataset comprises compounds sharing the same crystal structure: the half-Heusler compounds belong to space group 216, the spinels -- to space group 227, and the compounds from the dataset generated in this work -- to space group 225. This structural uniformity within each dataset makes all three datasets suitable for our SBFS-FT approach. The training sets provided with the HH130 and MDR datasets, and the configurations selected from the MD trajectories, were used to fine-tune the MatterSim potential.

\subsection{\label{sec:bte}Calculation of LTC with the approach based on the solution of Boltzmann transport equation}

A method based on the solution of Boltzmann transport equation (BTE) in the relaxation time approximation implemented in Phono3py package~\cite{ltc_bte, Togo2023_phono3py_implement, Togo2015_phono3py_theory} was utilized for obtaining LTC. The procedure of LTC calculation in this BTE-based approach was the following. Firstly, the unit cell of the compound under interest was relaxed with the FIRE method and Frechet cell filter until the forces dropped below 10$^{-5}$ eV/\AA using ASE~\cite{ase}. Secondly, the second- and third-order force constants were calculated using the supercell method ~\cite{supercell_method} with atomic displacements of 0.03~\AA, as implemented in Phono3py~\cite{Togo2023_phono3py_implement}. The supercell sizes used in these calculations are given in Table S1 in the Supplementary Material. The second-order force constants were then used for obtaining the phonon dispersion relations. From these phonon dispersions, the group velocities $\textbf{v}_{q\nu}$ and heat capacities $C_{q\nu} $ were computed with the following equations:

\begin{equation}
    \textbf{v}_{q\nu} = \frac{\partial\omega_{q\nu}}{\partial q}
    \label{eq:gv}
\end{equation}

\begin{equation}
   C_{q\nu} = k_\mathrm{B}
\left(\frac{\hbar\omega_{q\nu}}{k_\mathrm{B} T} \right)^2
\frac{\exp(\hbar\omega_{q\nu}/k_\mathrm{B}
T)}{[\exp(\hbar\omega_{q\nu}/k_\mathrm{B} T)-1]^2}
\label{eq:heat_cap}
\end{equation}

where $v_{q \nu}$, $C_{q \nu}$ and $\omega_{q \nu}$ are the group velocity, heat capacity and frequency of the phonon mode $\nu$ at the point $q$ of Brillouin zone, respectively.  

Finally, LTC was calculated according to the following formula~\cite{Togo2015_phono3py_theory}:

\begin{equation}
    \kappa_L = \frac{1}{NV_c}\sum_{q\nu}\tau_{q\nu} C_{q\nu}\textbf{v}_{q\nu}\otimes\textbf{v}_{q\nu}
    \label{eq:kappa}
\end{equation}
where $C_{q \nu}$, $v_{q \nu}$ and $\tau_{q \nu}$ are the heat capacity, group velocity and lifetime of the phonon mode $\nu$ at the point $q$ of Brillouin zone, respectively (all of the present phonon modes are included in the summation).
$V_c$ is the unit cell volume, $N$ is the number of unit cells in the supercell. The $q$-point mesh sizes used in these calculations are given in Table S1 in the Supplementary Material.

The phonon lifetimes $\tau_{q \nu}$ were computed from the imaginary part of the phonon self-energies. For this, the third-order force constants were calculated. The mathematical details on phonon lifetimes calculation can be found in~\cite{Togo2015_phono3py_theory, Togo2023_phono3py_implement}.

\subsection{\label{sec:msim_description}MatterSim machine-learned interatomic potential}

MatterSim is a universal machine-learned interatomic potential based on a deep equivariant message-passing neural network~\cite{mattersim}. It represents the local atomic environment of each atom using a graph neural network that processes interatomic distances and angles through rotationally equivariant operations, enabling accurate predictions of energies and forces regardless of the crystal symmetry. The model was pre-trained on an extensive dataset of several million DFT calculations, covering a broad range of chemical compositions and structures. Two versions are available: MatterSim 1M, smaller and faster model with 880 thousand parameters, and MatterSim 5M, a larger model for more accurate calculations with 4.5 million parameters. Thanks to its extensive pre-training, MatterSim provides accurate out-of-the-box predictions for virtually any inorganic compound; however, the PES may exhibit a softening effect that degrades the accuracy of its derivatives, thus requiring additional fine-tuning for precise LTC calculations.

\subsubsection{\label{sec:finetune_description}Methods of MatterSim fine-tuning}

{To circumvent the prohibitive computational and memory costs associated with full fine-tuning, we employ Parameter-Efficient Fine-Tuning (PEFT) methods. These approaches freeze the pre-trained weight matrices of the MatterSim model and introduce a minimal set of trainable parameters to model the weight update. We evaluate three distinct PEFT strategies, all utilizing a low adaptation rank $r \ll d$ (where $d$ is the feature dimension) to ensure computational efficiency.

\textbf{Low-Rank Adaptation (LoRA).} 
LoRA \cite{hu2022lora} approximates the weight update matrix as the product of two low-rank matrices. Instead of updating the full pre-trained weight matrix $W_0 \in \mathbb{R}^{d \times d}$, LoRA introduces trainable matrices to yield an adapted weight given by:
\begin{equation}
W' = W_0 + B A
\end{equation}
where $A \in \mathbb{R}^{r \times d}$ and $B \in \mathbb{R}^{d \times r}$ are low-rank matrices. Matrix $A$ is initialized with a Gaussian distribution and $B$ with zeros, ensuring the update is exactly zero at the start of training, thereby preserving the model's initial behavior.

\textbf{Weight-Decomposed Low-Rank Adaptation (DoRA).} 
DoRA \cite{liu2024dora} enhances LoRA by decoupling the pre-trained weights into magnitude and directional components. The pre-trained weight is decomposed such that the magnitude is represented by a vector and the direction by a matrix. DoRA makes the magnitude vector a trainable parameter and applies the low-rank update exclusively to the directional component. The adapted weight is formulated as:
\begin{equation}
W' = m \frac{V + B A}{\|V + B A\|_c}
\end{equation}
where $m$ is the trainable magnitude vector, $V$ is the frozen directional matrix, and the denominator denotes the column-wise vector norm. This decomposition improves learning capacity and training stability by allowing independent adjustment of magnitude and direction, closely mimicking the learning behavior of full fine-tuning without adding inference latency.

\textbf{Tucker Decomposition and Low Tensor Rank Adaptation (LoTR).} 
To achieve even greater parameter efficiency across deep architectures, LoTR extends the low-rank concept across multiple layers simultaneously by leveraging tensor decomposition. The mathematical foundation of this approach is the Tucker-2 decomposition, which factorizes a three-dimensional tensor into a smaller core tensor and two factor matrices. 

In the context of neural networks, we can view the sequence of weight updates across $L$ layers as a single three-dimensional tensor $\Delta \mathcal{W} \in \mathbb{R}^{d \times d \times L}$, where the first two modes correspond to the matrix dimensions and the third mode corresponds to the layer index $s$. Applying a Tucker-2 decomposition along the first two modes yields:
\begin{equation}
\Delta \mathcal{W} = \mathcal{G} \times_1 A \times_2 B
\end{equation}
where $\mathcal{G} \in \mathbb{R}^{r \times r \times L}$ is the core tensor, and $A \in \mathbb{R}^{d \times r}$, $B \in \mathbb{R}^{d \times r}$ are the shared factor matrices. In index notation, this tensor multiplication is explicitly expressed as:
\begin{equation}
\Delta W_{ijs} = \sum_{p=1}^r \sum_{q=1}^r A_{ip} \mathcal{G}_{pqs} B_{jq}
\end{equation}
Because the layer index $s$ only appears in the core tensor, we can slice $\mathcal{G}$ along the third mode to obtain a specific core matrix $G_s \in \mathbb{R}^{r \times r}$ for each layer $s$. This directly translates the global tensor decomposition into the layer-specific low-rank update formula:
\begin{equation}
\Delta W_s = A G_s B^\top
\end{equation}
Here, the factor matrices $A$ and $B$ are shared globally across all adapted layers, while $G_s$ is specific to the $s$-th layer. This shared tensor structure drastically reduces the total number of trainable parameters compared to standard LoRA. Specifically, the parameter count scales as $\mathcal{O}(r^2 L + 2dr)$ rather than $\mathcal{O}(dr L)$, effectively decoupling the model depth from the feature dimension and offering superior efficiency for deep models like MatterSim.

\textbf{Targeted Layer Adaptation Strategy.} 
In our implementation, we strategically restrict the application of all PEFT modules exclusively to the initial atom embedding layer and the final output projection layers of the MatterSim architecture. This targeted approach is motivated by two key factors. First, the intermediate message-passing layers of MatterSim already encode robust, universal physical laws and interatomic interactions learned from extensive pre-training on millions of density functional theory calculations. Freezing these intermediate layers prevents catastrophic forgetting of these fundamental physics representations. Second, the initial embedding and final projection layers are primarily responsible for mapping specific chemical inputs to the latent space and decoding latent representations into task-specific outputs, such as energies and forces. Adapting only these boundary layers drastically reduces the number of trainable parameters and optimizer memory footprint, while still providing sufficient capacity to align the pre-trained model with the symmetry-specific data distribution of our downstream tasks.}

\subsection{\label{sec:ltc_full_methodology}Methodology for accurate and efficient LTC calculation}

The methodology described in the present work allowed quickly and accurately compute thermal conductivities for large datasets of \textit{similar} compounds - compounds with the same crystal lattice symmetry. The first step of this methodology was to make a training subset from each dataset of \textit{similar} compounds. This subsets should consist of 20 compounds with various average atomic masses. The average atomic masses of the compounds in the subset should cover approximately the whole range of average atomic masses present in the initial \textit{similar} compounds dataset. The lists of compounds included in the training subsets from the HH130, 225, and spinels datasets are given in the Supplementary Material.

The configuration set for MatterSim fine-tuning was created by joining the sets of configurations (structures with displaced atoms or molecular dynamics (MD) trajectory frames) related to each compound from the subset. In the present work, for each compound these sets consisted of about 200-220 configurations. After the subset was created, MatterSim potential was fine-tuned for 700 epochs on this configurations set. This process of fine-tuning on the subset created as described above is referred to as SBFS-FT. Additionally, specific MatterSim potential was obtained for each compound by fine-tuning pre-trained MatterSim on the configurations related to this compound only -- this procedure is referred to as specific fine-tuning (SFT).

In order to test the performance and accuracy of our methodology, 8 other test compounds (not included in the subset for SBFS-FT) were selected. For each of the 28 compounds (20 compounds from the training subset and 8 test compounds), LTC was calculated using MTPs, MatterSim potentials created with SFT, and MatterSim after SBFS-FT. The LTC values were computed with the BTE-based approach for the temperatures of 300, 400, 500, 600, 700, 800, and 900 K. 
As a criterion of LTC prediction accuracy, the mean absolute relative differences $\varepsilon$ between the thermal conductivity values obtained with MTP and MatterSim were calculated with the following formula:
\begin{equation}
    \langle\varepsilon\rangle_{N_m} = \frac{1}{N_T\cdot N_m}\sum_{i=1}^{N_T}\sum_{j=1}^{N_m} \frac{|\kappa_0^{i,j}-\kappa^{i,j}|}{\kappa_0^{i,j}}\cdot100\%.
\end{equation}
Here $N_m$ is the number of compounds in the test set, $N_T$ is the number of temperature points $T_i$, $\kappa_0^{i,j}$ and $\kappa^{i,j}$ are LTC values for the compound $j$ at temperature $T_i$ with MTP and MatterSim, respectively.

\section{\label{sec:res-dis}Results and discussion}

\subsection{\label{sec:FT-method}The fine-tuning methods comparison}

Before applying the SBFS-FT methodology to compute the LTC, it is necessary to determine the most effective parameter-efficient fine-tuning (PEFT) strategy for MatterSim. While full fine-tuning of all model parameters yields high accuracy, it is computationally prohibitive for high-throughput screening. Consequently, we evaluate several advanced PEFT techniques to identify the optimal balance between predictive accuracy and computational overhead. Since LTC calculations via the BTE-based approach are based on the computation of phonon properties, we first benchmark these adaptation methods on a fast auxiliary task: the prediction of phonon density of states (DOS). The results of this comparative analysis, which ultimately guided the selection of the fine-tuning strategy for our subsequent high-throughput LTC calculations, are detailed in the next section.

To determine the most effective parameter-efficient fine-tuning (PEFT) strategy for our approach, we compared various adaptation methods on an auxiliary task of phonon density of states (DOS) prediction. The accuracy of the predicted phonon DOS was evaluated using the Tanimoto similarity index between the reference (MTP-based) and predicted DOS for a representative set of eight test compounds. The results of this comparison are summarized in Table~\ref{tab:materials_3digits}.
As anticipated, the universal MatterSim 1M model without fine-tuning (M1m) yields the lowest accuracy, with an average Tanimoto score of 0.595. Applying a naive fine-tuning approach to the target layers improves the average score to 0.775. Among the evaluated PEFT methods, DoRA demonstrates a lower performance (0.680) than the naive approach, suggesting that the magnitude-direction decoupling inherent to DoRA may not be optimal for adapting the specific physical representations encoded in MatterSim. In contrast, LoRA significantly enhances the prediction accuracy, achieving an average score of 0.835. The LoTR method yields the best overall results across the tested compounds, reaching the highest average Tanimoto index of 0.843. By leveraging tensor decomposition to share factors across layers, LoTR provides a highly efficient and expressive adaptation for the deep architecture of MatterSim. Given its superior performance in capturing the essential phonon features, the LoTR method was selected for the subsequent symmetry-based few-shot fine-tuning (SBFS-FT) and high-throughput LTC calculations.

\begin{table}[htbp]
\centering
\caption{Tanimoto similarity scores for the phonon density of states predicted by MatterSim 1M fine-tuned using different methods. The scores are evaluated over eight test compounds. `M1m' denotes the universal model without fine-tuning. `Avg.' represents the average score across all methods for each compound (rightmost column) and the average score of each method across all compounds (bottom row).}
\label{tab:materials_3digits}
\begin{ruledtabular}
\begin{tabular}{lccccc|c}
{Compound} & {Naive} & {DoRA} & {LoRA} & {LoTR} & {M1m} & {Avg.} \\
\hline
LiCaAs   & 0.609 & 0.667 & 0.633 & 0.653 & 0.612 & 0.635 \\
PdCeBi   & 0.744 & 0.740 & 0.893 & 0.888 & 0.715 & 0.796 \\
CoTeLa   & 0.747 & 0.709 & 0.813 & 0.827 & 0.679 & 0.755 \\
TiCoSb   & 0.919 & 0.906 & 0.878 & 0.885 & 0.450 & 0.808 \\
LiNZn    & 0.870 & 0.722 & 0.964 & 0.975 & 0.822 & 0.871 \\
LiSiSc   & 0.850 & 0.454 & 0.926 & 0.941 & 0.563 & 0.747 \\
TiTeOs   & 0.596 & 0.540 & 0.663 & 0.657 & 0.437 & 0.579 \\
SnPtTh   & 0.863 & 0.699 & 0.912 & 0.916 & 0.485 & 0.775 \\
\hline
{Avg.} & {0.775} & {0.680} & {0.835} & \textbf{0.843} & {0.595} & {0.746} \\
\end{tabular}
\end{ruledtabular}
\end{table}



\subsection{\label{sec:hh130-res} Effect of fine-tuning on LTC calculations accuracy}

We began testing the accuracy and performance of MatterSim in LTC calculations using the HH130 dataset. Both MatterSim 1M and 5M models were employed. After completing the SFT and SBFS-FT processes, we evaluated the accuracy of MatterSim by comparing the phonon DOS obtained based on the force constants computed with MTP and MatterSim (without fine-tuning, after SFT, and after SBFS-FT). Tanimoto similarity scores between the MTP‑ and MatterSim‑based DOS served as the accuracy metric. As shown in Tables \ref{tab:dos_tanimoto_test_msim1m} and \ref{tab:dos_tanimoto_test_msim5m}, for most test compounds, the Tanimoto scores for DOS obtained with MatterSim after SFT and after SBFS‑FT are close to each other and higher than those computed using non‑fine‑tuned MatterSim. Visual comparison of the DOS graphs (see the Supplementary Material) also confirms that the densities of states obtained after fine‑tuning are closer to the MTP‑computed data than those from the non‑fine‑tuned MatterSim. Thus, the phonon DOS comparison indicates that SBFS‑FT improved the accuracy of second‑order force constants computation with both MatterSim 1M and 5M potentials in comparison to the non‑fine‑tuned models, achieving a level similar to that after SFT.

\begin{table}[t!]
\caption{\label{tab:dos_tanimoto_test_msim1m}%
Tanimoto similarity scores for the densities of states obtained with MTP and MatterSim 1M without fine-tuning ($\TS_{\WFT}$), MatterSim after SFT and SBFS-FT for the HH130 dataset ($\TS_{\SFT}$).
}
\begin{ruledtabular}
\begin{tabular}{cccc}
compound & $\TS_{\WFT}$ & $\TS_{\SFT}$ & $\TS_{\SBFT}$ \\ 
\colrule

LiNZn & 0.82 & 0.90 & 0.97 \\ 
CoSbTi & 0.45 & 0.85 & 0.88 \\ 
BiCePd & 0.72 & 0.79 & 0.89 \\ 
CoLaTe & 0.68 & 0.84 & 0.83 \\ 
LiScSi & 0.52 & 0.55 & 0.94 \\ 
AsCaLi & 0.61 & 0.92 & 0.65 \\ 
PtSnTh & 0.48 & 0.91 & 0.92 \\ 
OsTeTi & 0.44 & 0.71 & 0.66 \\ 

\end{tabular}
\end{ruledtabular}
\end{table}

\begin{table}[t!]
\caption{\label{tab:dos_tanimoto_test_msim5m}%
Tanimoto similarity scores for the densities of states obtained with MTP and MatterSim 5M without fine-tuning ($\TS_{\WFT}$), MatterSim after SFT and SBFS-FT for the HH130 dataset ($\TS_{\SFT}$).
}
\begin{ruledtabular}
\begin{tabular}{cccc}
compound & $\TS_{\WFT}$ & $\TS_{\SFT}$ & $\TS_{\SBFT}$ \\ 
\colrule

LiNZn & 0.40 & 0.81 & 0.98 \\ 
CoSbTi & 0.68 & 0.81 & 0.86 \\ 
BiCePd & 0.63 & 0.63 & 0.90 \\ 
CoLaTe & 0.64 & 0.86 & 0.75 \\ 
LiScSi & 0.72 & 0.57 & 0.60 \\
AsCaLi & 0.84 & 0.96 & 0.81 \\ 
PtSnTh & 0.61 & 0.97 & 0.90 \\ 
OsTeTi & 0.22 & 0.45 & 0.48 \\ 

\end{tabular}
\end{ruledtabular}
\end{table}

\begin{figure}[h!]
\includegraphics[width=1\linewidth]{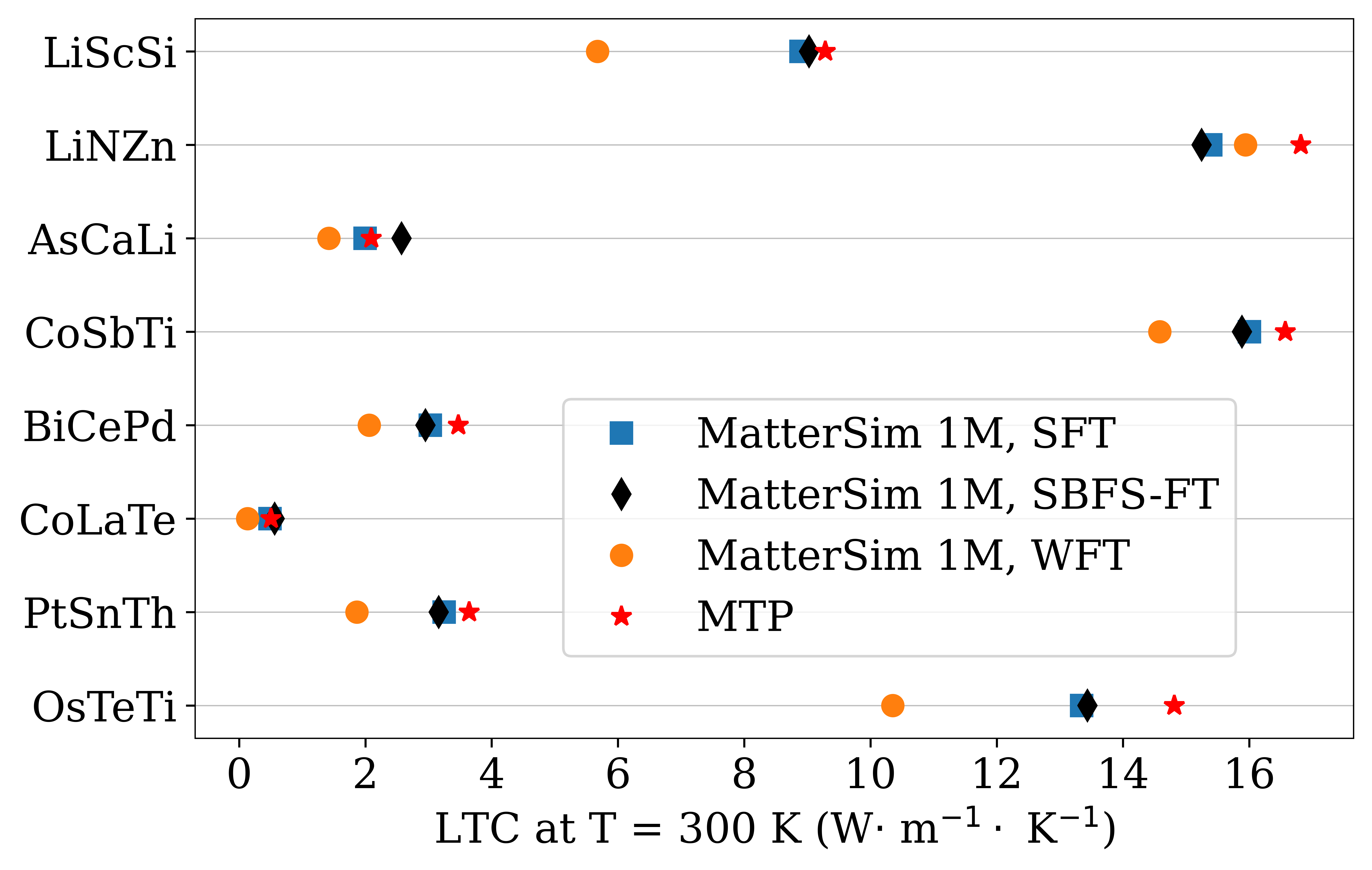}
\caption{\label{fig:msim1m_300K} Lattice thermal conductivities obtained with MTP, MatterSim 1M without fine-tuning (WFT), MatterSim 1M fine-tuned specifically for each test compound from the HH130 dataset (specific fine-tuning, SFT) and MatterSim 1M fine-tuned on the 20 compounds from HH130 dataset (SBFS-FT).}
\end{figure}

\begin{figure}[h!]
\includegraphics[width=1\linewidth]{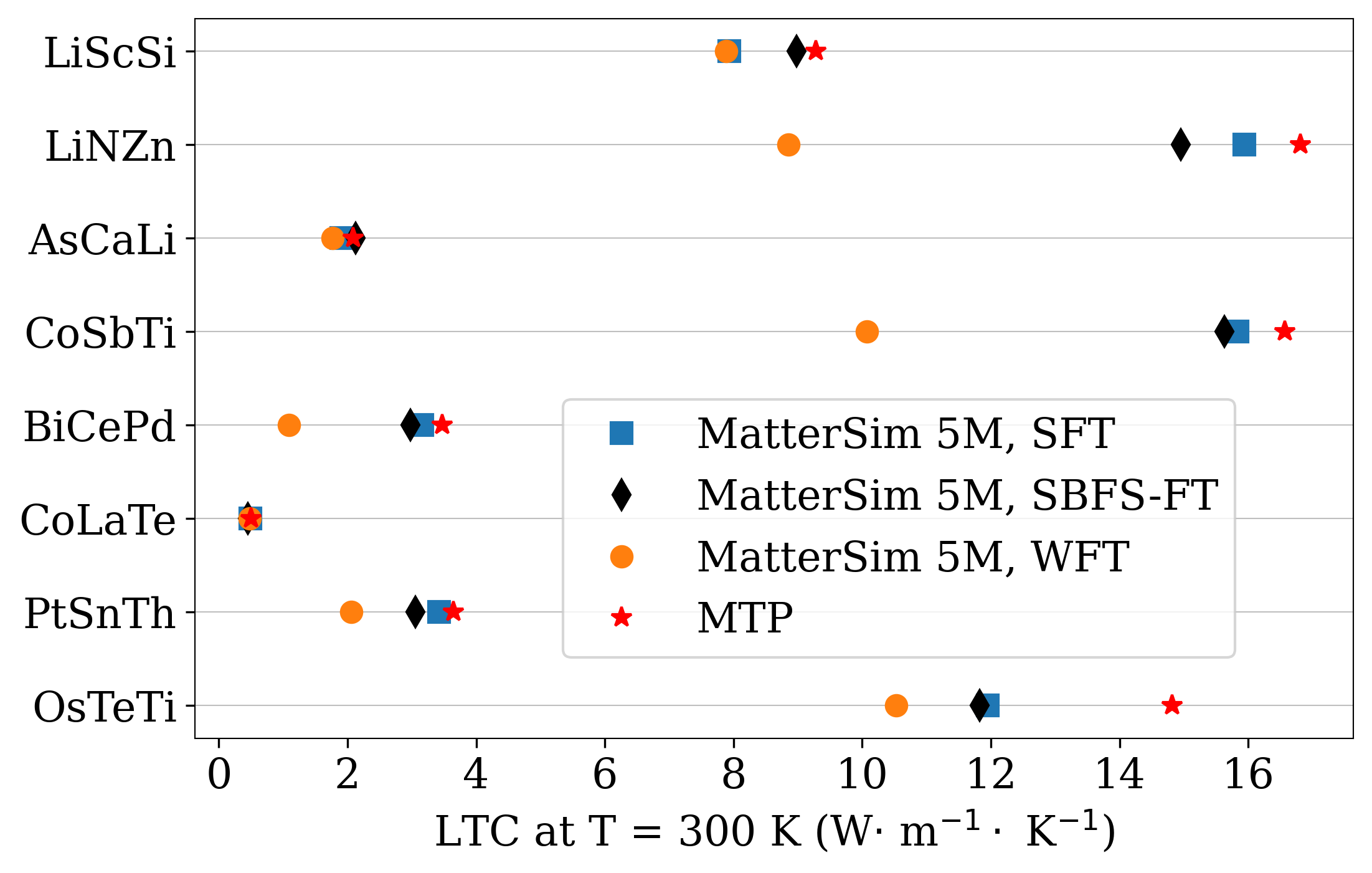}
\caption{\label{fig:msim5m_300K} Lattice thermal conductivities obtained with MTP, MatterSim 5M without fine-tuning, MatterSim 5M fine-tuned specifically for each test compound from the HH130 dataset (specific fine-tuning) and MatterSim 5M fine-tuned on the 20 compounds from HH130 dataset (SBFS-FT).}
\end{figure}

After DOS comparison, LTC values were calculated with MTP and the MatterSim models for the 28 compounds from the HH130 dataset. The resulting values of thermal conductivity at 300 K for the test compounds are shown in Figs. \ref{fig:msim1m_300K}, \ref{fig:msim5m_300K}. The mean absolute relative differences between the LTC values obtained with MTP and with MatterSim for the test compounds are presented in Table \ref{tab:msim_ltc_rel_diff}. Similar data for the training subset compounds can be found in the Supplementary Material. According to this data, SBFS-FT of MatterSim reduces the relative difference between MTP and MatterSim by about 3 times. Moreover, the accuracies of MatterSim models after SFT and SBFS-FT were similar ($\langle\varepsilon_{\SBFT}\rangle_8-\langle\varepsilon_{\SFT}\rangle_8 \sim 2-3 \%$). 

\begin{table}[t!]
\caption{\label{tab:msim_ltc_rel_diff}%
Mean absolute relative differences between the lattice thermal conductivities obtained with MTP and MatterSim 1M (5M) without fine-tuning ($\varepsilon_{\WFT}$), MatterSim 1M (5M) fine-tuned on the 20 compounds from HH130 dataset ($\varepsilon_{\SBFT}$) and MatterSim 1M (5M) fine-tuned specifically for each test compound from the HH130 dataset ($\varepsilon_{\SFT}$). Averaging $\langle \cdot \rangle_8$ was done over the 8 test compounds from the HH130 dataset and all temperature points (T = 300, 400, 500, 600, 700, 800, 900 K).
}
\begin{ruledtabular}
\begin{tabular}{cccc}
Model & $\langle\varepsilon_{\WFT}\rangle_8$, \%  & $\langle\varepsilon_{\SBFT}\rangle_8$, \%  & $\langle\varepsilon_{\SFT}\rangle_8$, \% \\ 
\colrule
 MatterSim 1M & 35.0 & 10.8 & 7.1 \\ 
MatterSim 5M & 32.5 & 10.3 & 8.8  \\ 

\end{tabular}
\end{ruledtabular}
\end{table}

Therefore, in the case of HH130 dataset, MatterSim potential trained by means of SBFS-FT allows calculating LTC with the accuracy close to that of the MatterSim potentials fine-tuned for each compound individually (even for the compounds not present in the training subset). 

In addition, we compared the accuracy and performance of MatterSim 1M and 5M models in computing thermal conductivity. According to the data given in Table \ref{tab:msim_ltc_rel_diff}, the accuracies of LTC calculations with MatterSim 1M and 5M are similar (the values $\langle\varepsilon_{\SBFT}\rangle_8$ are very close for these two models). However, the time taken by both SBFS-FT on a single NVIDIA A100 GPU~\cite{a100gpu} for the 28 compounds from the HH130 dataset was about 2 times less for MatterSim 1M than that for MatterSim 5M. Therefore, we have chosen MatterSim 1M for further LTC calculations in order to reduce the computational costs and time of fine-tuning.

\begin{table}[h]
\centering
\caption{Time taken by SFT and SBFS-FT of MatterSim on the HH130 dataset. Time taken by specific fine-tuning is the time required for preparing 28 potentials fine-tuned for each of the 28 compounds from the training and test subsets from the HH130 dataset on a single GPU. Time taken by SBFS-FT is the time required for fine-tuning MatterSim on the joined data for 20 compounds from the HH130 dataset (training subset).}
\label{tab:time_hh130}
\begin{ruledtabular}
\begin{tabular}{ccc}
Fine-tuning method & Model & GPU hours \\
\colrule
SFT  & MatterSim 1M & 17.3 \\
SFT & MatterSim 5M &  35.0 \\
SBFS-FT & MatterSim 1M & 10.4 \\
SBFS-FT & MatterSim 5M & 22.4 \\
\end{tabular}
\end{ruledtabular}
\end{table}

Similarly to the case of HH130 dataset, the use of SBFS-FT for improving the accuracy of LTC calculations was tested on the 225 and spinels datasets. Details on phonon DOS calculations and comparison for these dataset are given in the Supplementary Material. The relative differences between the LTC values obtained with MTP and MatterSim 1M without fine-tuning, after SBFS-FT and after SFT (averaged over the 8 test compounds) are given in Table~\ref{tab:rel_diff_spin_225}. Therefore, in the present case fine-tuning also led to the increase of LTC computation accuracy. Moreover, the accuracy of LTC calculations with the models obtained with SBFS-FT and SFT was also similar.

\begin{table}[t!]
\caption{\label{tab:rel_diff_spin_225}%
mean absolute relative differences between the lattice thermal conductivities obtained with MTP and MatterSim 1M without fine-tuning ($\varepsilon_{\WFT}$), MatterSim 1M fine-tuned with SBFS-FT ($\varepsilon_{\SBFT}$) and MatterSim 1M (5M) fine-tuned specifically for each test compound ($\varepsilon_{\SFT}$). Averaging $\langle \cdot \rangle_8$ was done over the 8 test compounds from the corresponding dataset and all temperature points (T = 300, 400, 500, 600, 700, 800, 900 K).
}
\begin{ruledtabular}
\begin{tabular}{cccc}
Dataset & $\langle\varepsilon_{\WFT}\rangle_8$, \%  & $\langle\varepsilon_{\SBFT}\rangle_8$, \%  & $\langle\varepsilon_{\SFT}\rangle_8$, \% \\ 
\colrule
225 dataset & 32.3 & 16.0 & 22.6 \\ 
Spinels & 21.1 & 10.8 & 21.6  \\ 

\end{tabular}
\end{ruledtabular}
\end{table}

\begin{figure}[h!]
\includegraphics[width=1\linewidth]{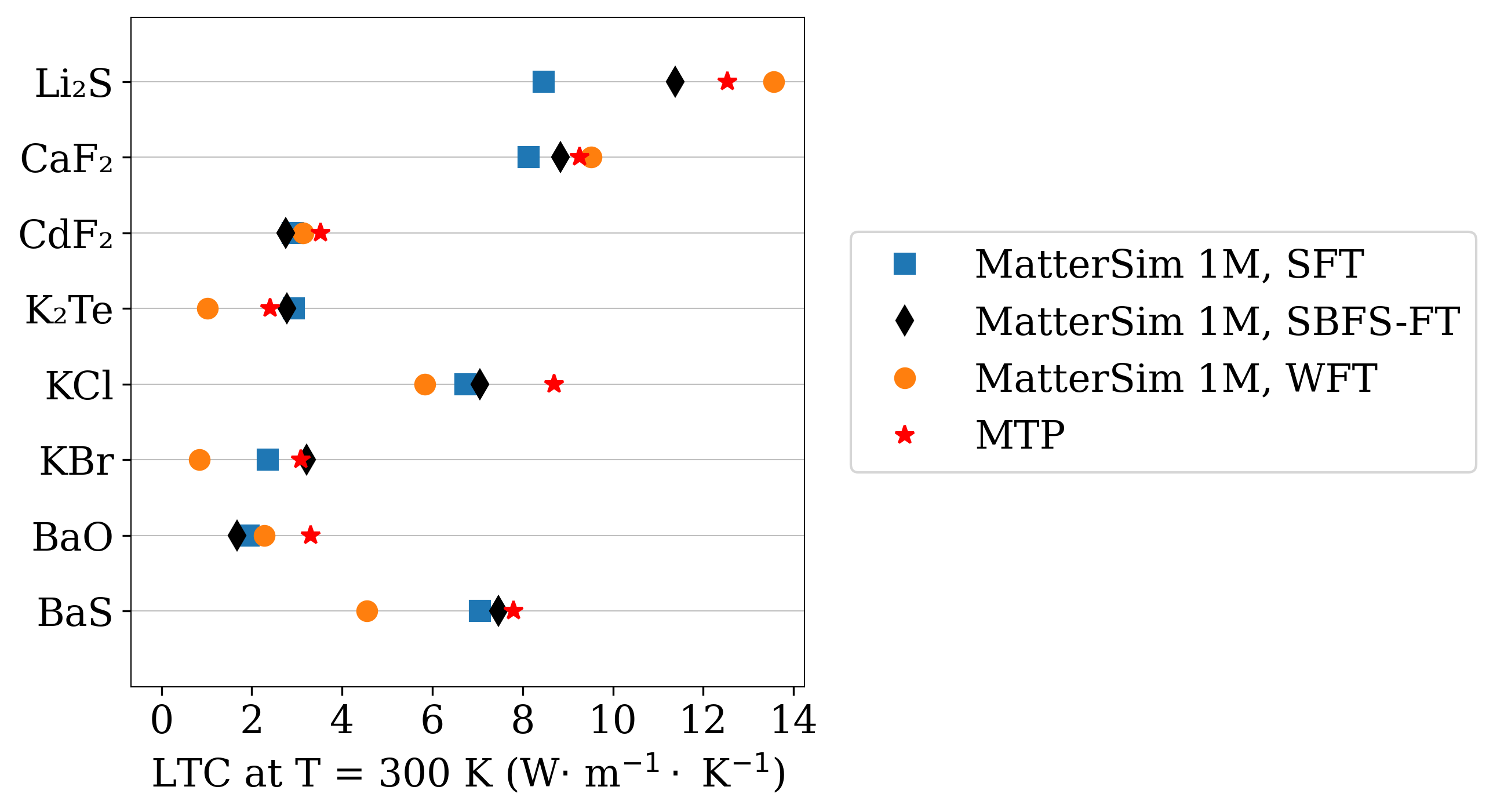}
\caption{\label{fig:msim1m_300K_225} Lattice thermal conductivities obtained with MTP, MatterSim 1M without fine-tuning (WFT), MatterSim 1M fine-tuned specifically for each test compound from the 225 dataset (specific fine-tuning, SFT) and MatterSim 1M fine-tuned on the 20 compounds from 225 dataset (SBFS-FT).}
\end{figure}

\begin{figure}[h!]
\includegraphics[width=1\linewidth]{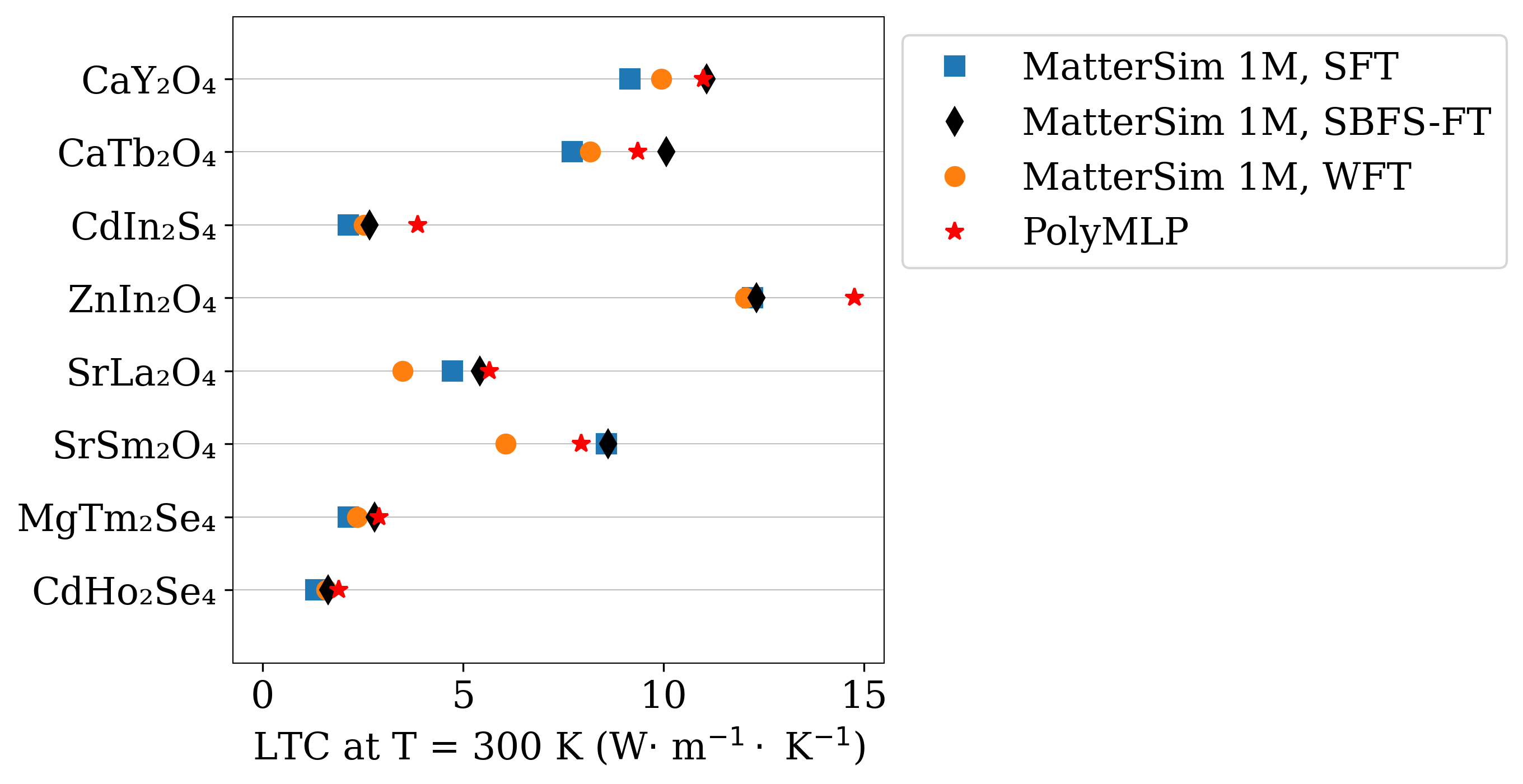}
\caption{\label{fig:msim1m_300K_225} Lattice thermal conductivities obtained with MTP, MatterSim 1M without fine-tuning (WFT), MatterSim 1M fine-tuned specifically for each test compound from the spinels dataset (specific fine-tuning, SFT) and MatterSim 1M fine-tuned on the 20 compounds from 225 dataset (SBFS-FT).}
\end{figure}

\subsection{Time efficiency comparison for SFT and SBFS-FT}

\begin{figure}[h!]
\includegraphics[width=\linewidth]{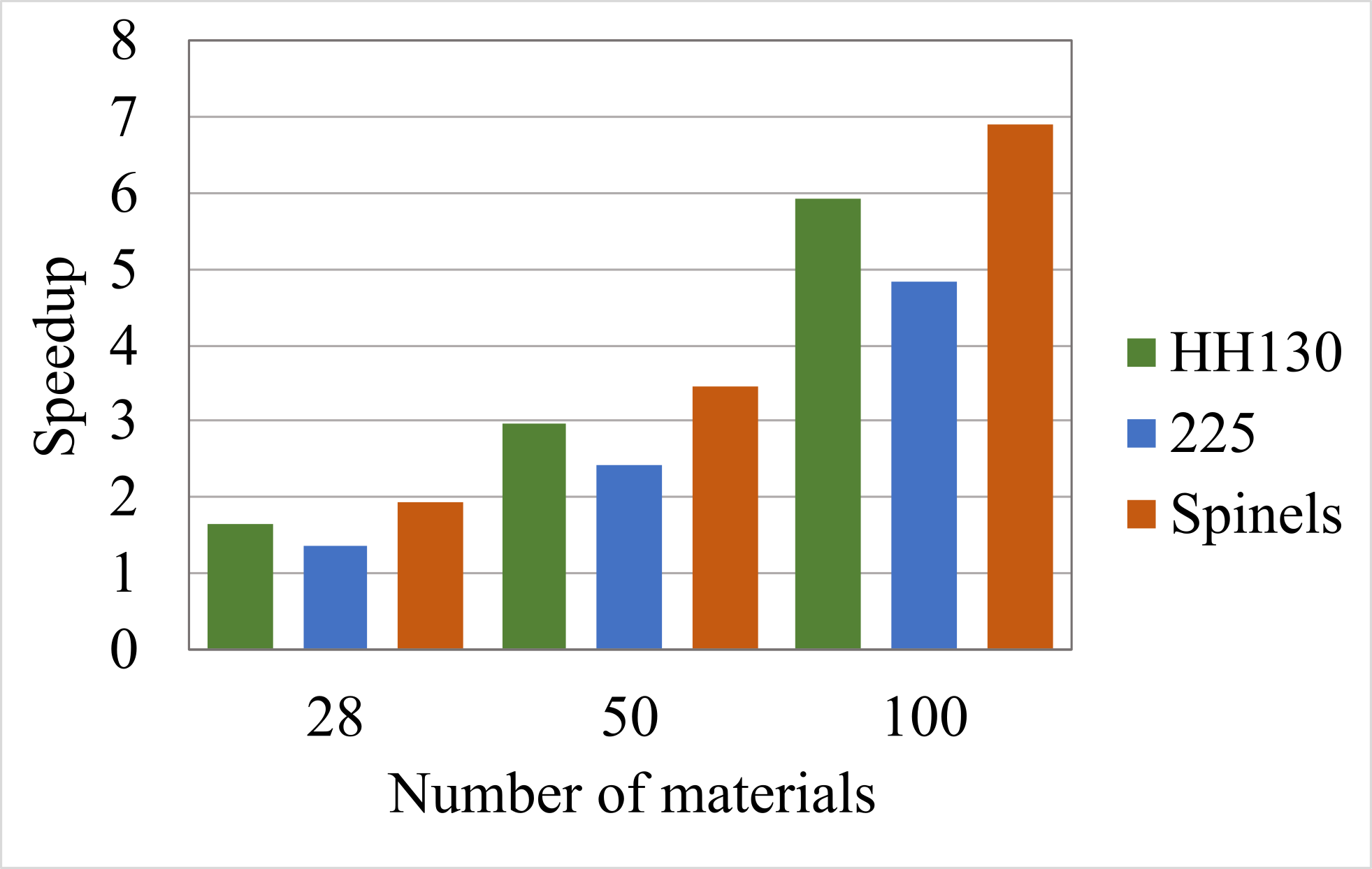}
\caption{\label{fig:speedup} Speedup of fine-tuning (ratio between the time taken by SFT for $N$ compounds and SBFS-FT on 20 compounds) for $N=28,50,100$ compounds from the HH130, 225, and spinels datasets. Here we compute the time taken by SFT as $N\cdot t_1$, where $N$ is the number of compounds and $t_1$ is the time required for fine-tuning MatterSim 1M on the set of about 220 configurations for a single compound. $t_1$ is about 1, 0.62, and 0.67 GPU hours for the 225, HH130, and spinels datasets, respectively.}
\end{figure}

After finding that SBFS-FT and SFT allow predicting LTC with approximately similar accuracy, we compared the time taken by these two fine-tuning procedures on a single NVIDIA A100 GPU. We found that for all three datasets, SBFS-FT was faster than SFT. The average time taken by fine-tuning on of MatterSim 1M on the set of about 220 configurations for 1 compound was  0.62, 1, and 0.67 GPU hours for the HH130, 225, and spinels  datasets, respectively. SBFS-FT on 20 compounds took 10.4, 20.6, and 9.7 GPU hours for HH130, 225 and spinels datasets, respectively. Therefore, even in the case of LTC calculations for 28 compounds performed in the present work, the time taken by SBFS-FT was about 1.5--2 times less than that taken by specific fine-tuning for all datasets. Speedup estimated for larger amount of compounds is given in Fig.\ \ref{fig:speedup}. 

\begin{figure}[h!]
\includegraphics[width=0.9\linewidth]{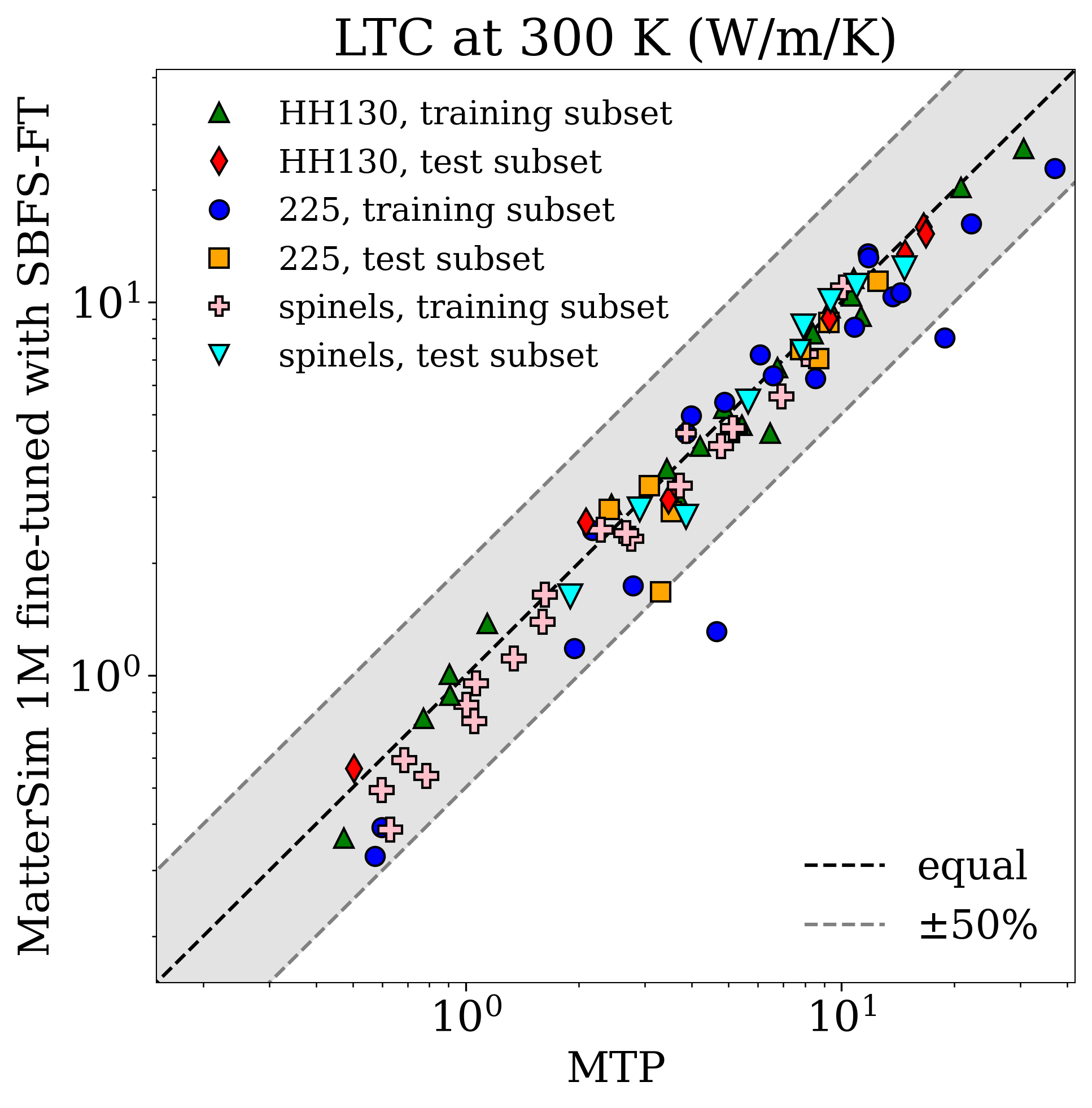}
\caption{\label{fig:ltc_all} Comparison of the LTC values computed at 300 K with the BTE-based approach using MTP and MatterSim after SBFS-FT for the 28 compounds from the HH130 dataset, the 28 compounds from the 225 dataset, and the 28 compounds from the spinels dataset.}
\end{figure}

To sum up, in the case of high-throughput LTC calculations for all three datasets, SBFS-FT and SFT have similar accuracy of LTC predictions, but  SBFS-FT is more computationally efficient than SFT. Comparison of the LTC values obtained with MTP and MatterSim after SBFS-FT for the compounds from all three datasets is shown in Fig.\ \ref{fig:ltc_all}. The LTC values computed using MatterSim after SBFS-FT are more than 2 times lower than those obtained with the MTPs for two compounds from the 225 dataset, namely PbTe and KH. The origin of this LTC discrepancy is discussed in the Supplementary Material.

\section{\label{sec:conclusions}Conclusions}

This study evaluated the effectiveness of symmetry-based few-shot fine-tuning (SBFS-FT) in improving the accuracy of high-throughput LTC calculations using the MatterSim potential. The half-Heusler compounds (HH130) dataset, a set of compounds with space group 225, and a dataset of spinels from the MDR database were used for testing our methodology.

The results demonstrate that SBFS-FT significantly enhances the accuracy of second-order force constants calculations, as evidenced by higher Tanimoto similarity scores for phonon DOS after fine-tuning compared to the non-fine-tuned MatterSim. For thermal conductivity calculations, SBFS-FT reduced the mean absolute relative difference between MTP and MatterSim LTC values in comparison to non-fine-tuned models, achieving accuracies comparable to those obtained with specific fine-tuning of MatterSim. Notably, this improved accuracy extended to compounds not present in the training subset, indicating good generalization. Additionally, both MatterSim 1M and 5M models achieved similar LTC prediction accuracies after fine-tuning. However, MatterSim 1M required about half the computational time for SBFS‑FT and SFT as MatterSim 5M, making it a more efficient choice for high‑throughput LTC calculations.

Crucially, SBFS-FT consistently outperformed SFT in computational efficiency across all three datasets. The time taken by SBFS-FT was approximately 1.5-2 times less than that taken by SFT. This efficiency gain becomes more pronounced with larger compound sets, as SBFS-FT requires only a single fine-tuning run, whereas SFT necessitates separate fine-tuning for each compound. For 100 compounds, the projected speedup is approximately 5-6 times for each dataset.

In summary, SBFS-FT of MatterSim 1M offers an optimal balance between accuracy and computational cost for high-throughput LTC calculations. It helps to achieve an accuracy level close to that of SFT while substantially reducing fine-tuning time, making it a practical and scalable approach for accelerating LTC predictions across large materials datasets. The strong agreement between MTP and MatterSim after SBFS-FT (Fig.~\ref{fig:ltc_all}) further validates this method as a reliable alternative to costly \textit{ab initio} methods and compound-specific fine-tuning in large-scale materials screening workflows.

\bibliography{bibliography}

@PREAMBLE{
 "\providecommand{\noopsort}[1]{}" 
 # "\providecommand{\singleletter}[1]{#1}%" 
}

@article{mtp_proofs_1,
  title={Accessing thermal conductivity of complex compounds by machine learning interatomic potentials},
  author={Korotaev, Pavel and Novoselov, Ivan and Yanilkin, Aleksey and Shapeev, Alexander},
  journal={Physical Review B},
  volume={100},
  number={14},
  pages={144308},
  year={2019},
  publisher={APS}
}

@article{mtp_proofs_2,
  title={Electronic and thermal transport in novel carbon-based bilayer with tetragonal rings: a combined study using first-principles and machine learning approach},
  author={Ghosal, Supriya and Chowdhury, Suman and Jana, Debnarayan},
  journal={Physical Chemistry Chemical Physics},
  volume={23},
  number={27},
  pages={14608--14616},
  year={2021},
  publisher={Royal Society of Chemistry}
}

@article{mtp_proofs_3,
  title={Ultrahigh thermal conductivity and strength in direct-gap semiconducting graphene-like {BC}$_6${N}: A first-principles and classical investigation},
  author={Mortazavi, Bohayra},
  journal={Carbon},
  volume={182},
  pages={373--383},
  year={2021},
  publisher={Elsevier}
}

@article{mtp_proofs_4,
  title={A moment tensor potential for lattice thermal conductivity calculations of $\alpha$ and $\beta$ phases of Ga2O3},
  author={Rybin, Nikita and Shapeev, Alexander},
  journal={Journal of Applied Physics},
  volume={135},
  number={20},
  year={2024},
  publisher={AIP Publishing}
}

@article{softening,
  title={Systematic softening in universal machine learning interatomic potentials},
  author={Deng, Bowen and Choi, Yunyeong and Zhong, Peichen and Riebesell, Janosh and Anand, Shashwat and Li, Zhuohan and Jun, KyuJung and Persson, Kristin A and Ceder, Gerbrand},
  journal={npj Computational Materials},
  volume={11},
  number={1},
  pages={9},
  year={2025},
  publisher={Nature Publishing Group UK London}
}

@article{hh130,
  title={HH130: a standardized database of machine learning interatomic potentials, datasets, and its applications in the thermal transport of half-Heusler thermoelectrics},
  author={Yang, Yuyan and Lin, Yifei and Dai, Shengnan and Zhu, Yifan and Xi, Jinyang and Xi, Lili and Gu, Xiaokun and Singh, David J and Zhang, Wenqing and Yang, Jiong},
  journal={Digital Discovery},
  volume={3},
  number={11},
  pages={2201--2210},
  year={2024},
  publisher={Royal Society of Chemistry}
}

@article{mattersim,
  title={Mattersim: A deep learning atomistic model across elements, temperatures and pressures},
  author={Yang, Han and Hu, Chenxi and Zhou, Yichi and Liu, Xixian and Shi, Yu and Li, Jielan and Li, Guanzhi and Chen, Zekun and Chen, Shuizhou and Zeni, Claudio and others},
  journal={arXiv preprint arXiv:2405.04967},
  year={2024}
}

@article{ltc_bte,
  title = {Lattice thermal conductivities of two {SiO}$_2$ polymorphs by first-principles calculations and the phonon Boltzmann transport equation},
  author = {Mizokami, Keiyu and Togo, Atsushi and Tanaka, Isao},
  journal = {Phys. Rev. B},
  volume = {97},
  issue = {22},
  pages = {224306},
  numpages = {10},
  year = {2018},
  month = {Jun},
  publisher = {American Physical Society}
}

@article{Togo2015_phono3py_theory,
  title = {Distributions of phonon lifetimes in Brillouin zones},
  author = {Togo, Atsushi and Chaput, Laurent and Tanaka, Isao},
  journal = {Phys. Rev. B},
  volume = {91},
  issue = {9},
  pages = {094306},
  numpages = {31},
  year = {2015},
  month = {Mar},
  publisher = {American Physical Society}
}

@article{Togo2023_phono3py_implement,
year = {2023},
month = {jun},
publisher = {IOP Publishing},
volume = {35},
number = {35},
pages = {353001},
author = {Atsushi Togo and Laurent Chaput and Terumasa Tadano and Isao Tanaka},
title = {Implementation strategies in phonopy and phono3py},
journal = {Journal of Physics: Condensed Matter},
}

@article{supercell_method,
  title = {First-Principles Determination of the Soft Mode in Cubic {${\mathrm{ZrO}}_{2}$}},
  author = {Parlinski, K. and Li, Z. Q. and Kawazoe, Y.},
  journal = {Physical Review Letters},
  volume = {78},
  issue = {21},
  pages = {4063--4066},
  numpages = {0},
  year = {1997},
  month = {May},
  publisher = {American Physical Society}
}

@article{fhiaims,
  title={Ab initio molecular simulations with numeric atom-centered orbitals},
  author={Blum, Volker and Gehrke, Ralf and Hanke, Felix and Havu, Paula and Havu, Ville and Ren, Xinguo and Reuter, Karsten and Scheffler, Matthias},
  journal={Computer Physics Communications},
  volume={180},
  number={11},
  pages={2175--2196},
  year={2009},
  publisher={Elsevier}
}

@article{a100gpu,
  title={Nvidia a100 tensor core gpu: Performance and innovation},
  author={Choquette, Jack and Gandhi, Wishwesh and Giroux, Olivier and Stam, Nick and Krashinsky, Ronny},
  journal={IEEE Micro},
  volume={41},
  number={2},
  pages={29--35},
  year={2021},
  publisher={IEEE}
}

@article{ase,
  title={The atomic simulation environment—a Python library for working with atoms},
  author={Hjorth Larsen, Ask and J{\o}rgen Mortensen, Jens and Blomqvist, Jakob and Castelli, Ivano E and Christensen, Rune and Du{\l}ak, Marcin and Friis, Jesper and Groves, Michael N and Hammer, Bj{\o}rk and Hargus, Cory and others},
  journal={Journal of Physics: Condensed Matter},
  volume={29},
  number={27},
  pages={273002},
  year={2017},
  publisher={IOP Publishing}
}

@article{shapeev2016moment,
  title={Moment tensor potentials: A class of systematically improvable interatomic potentials},
  author={Shapeev, Alexander V},
  journal={Multiscale Modeling \& Simulation},
  volume={14},
  number={3},
  pages={1153--1173},
  year={2016},
  publisher={SIAM}
}

@inproceedings{
hu2022lora,
title={Lo{RA}: Low-Rank Adaptation of Large Language Models},
author={Edward J Hu and yelong shen and Phillip Wallis and Zeyuan Allen-Zhu and Yuanzhi Li and Shean Wang and Lu Wang and Weizhu Chen},
booktitle={International Conference on Learning Representations},
year={2022},
url={https://openreview.net/forum?id=nZeVKeeFYf9}
}

@inproceedings{
liu2024dora,
title={Do{RA}: Weight-Decomposed Low-Rank Adaptation},
author={Shih-yang Liu and Chien-Yi Wang and Hongxu Yin and Pavlo Molchanov and Yu-Chiang Frank Wang and Kwang-Ting Cheng and Min-Hung Chen},
booktitle={Forty-first International Conference on Machine Learning},
year={2024},
url={https://openreview.net/forum?id=3d5CIRG1n2}
}

@article{togo2023mdr,
  title={MDR phonon calculation database},
  author={Togo, Atsushi},
  journal={National Institute for Materials Science (NIMS)},
  year={2023}
}

@article{polymlp_repository,
    author = {Seko, Atsuto},
    title = {Tutorial: Systematic development of polynomial machine learning potentials for elemental and alloy systems},
    journal = {J. Appl. Phys.},
    volume = {133},
    number = {1},
    pages = {011101},
    year = {2023},
    month = {01},
}

@article{polymlp_ltc,
  title={On-the-fly training of polynomial machine learning potentials in computing lattice thermal conductivity},
  author={Togo, Atsushi and Seko, Atsuto},
  journal={The Journal of Chemical Physics},
  volume={160},
  number={21},
  year={2024},
  publisher={AIP Publishing}
}

\end{document}